\documentclass[prc,aps,amsfonts,twocolumn,dvips]{revtex4}
\usepackage{graphics,epsfig,amssymb}

\begin{document}

\title{\bf
Convergence of Particle-Hole Expansions for the Description of
Nuclear Correlations
}

\author{\rm N. Pillet$^{(a,c)}$,
N. Sandulescu$^{(b,c)}$, Nguyen Van Giai$^{(c)}$ and J.-F. Berger$^{(a)}$ }

\bigskip

\affiliation{\rm
$^{(a)}$Service de Physique Nucl\'eaire, CEA/DAM-Ile de France, BP12,
F-91680 Bruy\`eres-le-Ch\^atel, France \\
$^{(b)}$Institute of Physics and Nuclear Engineering, 76900 Bucharest,
Romania \\
$^{(c)}$Institut de Physique Nucl\'eaire, Universit\'e Paris-Sud,
F-91406 Orsay Cedex, France }

\date{\today}

\def\fid{\vert\phi >}
\def\fig{< \phi\vert}
\def\psid{\vert\Psi>}
\def\psig{<\Psi\vert}
\def\psid{\vert\Psi>}
\def\psig{<\Psi\vert}
\def\dspt{\displaystyle}

\begin{abstract}
The convergence properties of a multiparticle-multihole (m$p$-m$h$)
configuration mixing approach whose purpose is to describe ground state
correlations in nuclei without particle number and Pauli violations is
investigated in the case of an exactly solvable pairing hamiltonian.
Two different truncation schemes are tested by looking at quantities
as correlation energies and single-particle occupation probabilities.
Results show that pairing correlations present in usual superfluid nuclei
can be accurately described using up to 6 particle--6 hole excitations,
a convergence fast enough for envisaging extensions to
fully microscopic calculations.

\end{abstract}

\maketitle

\section{Introduction}

One of the most powerful method of predicting nuclear ground state properties
and excitations all over the nuclear chart is the microscopic approach based on
the self-consistent mean-field theory~\cite{rs}. Within this kind of approach,
the nuclear mean-field is determined from a variational technique of the
Hartree-Fock (HF) type applied to a many-body Hamiltonian. In order to be
tractable up to the heaviest nuclei, the method is usually applied by
employing phenomenological effective two-body
interactions such as
the Skyrme~\cite{skyrme} or the Gogny~\cite{gogny} forces.

It is well known that an accurate description of nuclear structure almost
always requires taking into account correlations beyond the simple HF
approximation. One important class of these correlations, namely pairing
correlations, are commonly
treated using the BCS or Hartree-Fock-Bogoliubov (HFB) theories. A second class
of correlations comes from the coupling of the nucleon motion to collective
oscillations of the mean field. Depending on the amplitude of collective
oscillations, these correlations are usually derived either from the Random
Phase Approximation (RPA) theory or from the Generator Coordinate Method (GCM).
Let us emphasize that, in addition to numerous standard nuclear properties,
correlations are expected to play a key role in the interpretation of the
unusual structure recently found in some nuclei such as the so-called parity
inversions observed in the ground state and low-lying excitations of
exotic light nuclei~\cite{exotic1,exotic}.

Most techniques used for including correlations suffer from defects that may be
an obstacle to an accurate description of
nuclear structure observables.
For instance, proton and neutron numbers are not conserved in the BCS
and HFB schemes. This deficiency is known to render these theories inadequate
for describing superfluid to normal phase transitions and, more generally,
situations where pairing correlations
are small.
As a consequence, the precision of the approximations consisting in treating
simple excitations as multi-quasiparticle states in even-even and odd-even nuclei
becomes questionable.
Similarly, the quasi-boson approximation employed in RPA generates violations
of the Pauli principle that require to introduce corrections to
the mean-value of one-body operators and render delicate a microscopic
treatment of particle-vibration coupling.

These difficulties can be obviated by having recourse to particle number
projection techniques in the case of pairing or to extensions of RPA
that do not make use of
the quasi-boson approximation. However, these techniques
complicate numerical algorithms considerably. For instance, in order to get a
realistic description of weak pairing situations necessitates to perform the
particle number projections {\it before} variation~\cite{projection}.
In the same way, extensions
of the RPA avoiding the violation of Pauli correlations as e.g., the
Self-Consistent RPA~\cite{scrpa} require significant
additional numerical effort.

In view of this, an alternative approach has been proposed in
Ref.\cite{Pillet02}, which consists of taking a ground state trial wave
function in the form of a linear combination of multiparticle-multihole
(m$p$-m$h$) operators acting upon the HF state.
The relevance of taking as a starting point a set of HF
single-particle states has been pointed out in Ref.\cite{Brown}.
The m$p$-m$h$ expansion is thought to be truncated at
a given order and the mixing coefficients are determined by minimizing the
total energy of the system.
The interest of such a trial wave function is of course to exactly conserve
nucleon numbers and be fully consistent with the Pauli principle.
In lowest order, in which only $1p$-$1h$
configurations are considered, one gets the well-known Tamm-Dancoff
approximation (TDA) \cite{rs}. Adding $2p$-$2h$ and higher order configurations
allows one to describe ground state correlations of the most general form,
including pairing and RPA correlations.
In the simplest application of this method, a HF calculation is performed and
the secular equation
involving the matrix elements of the residual interaction is diagonalized.
The coefficients of the m$p$-m$h$ superposition for the ground state are then
taken from the eigenvector belonging to the lowest eigenvalue. The
eigenvectors belonging to the next few higher eigenvalues
can be interpreted as approximately representing the lowest
excitations of the system.
Let us point out that the advantage of building  m$p$-m$h$ configurations
from HF single-particle states lies in the fact that an important
part of two-body correlations is already included in the nuclear mean field.
One therefore expects that the
effect of the residual interaction is small enough to allow a
fast convergence of the m$p$-m$h$ expansion
so that
only the lowest orders will have to be retained.

A more elaborate version of this method is to allow the HF single particle
themselves to be included as variational parameters in the energy minimization.
This procedure yields additional equations from which renormalized single
particle states can be derived. The method then is similar to the
Multi-Configuration Hartree-Fock (MCHF) approach which is widely used in atomic
and molecular physics~\cite{Froese69}.
The interest of introducing such a consistency between the single-particle
structure and the correlated ground state is that additional
parts of two-body correlations are further included in the nuclear mean field.
In this situation even lower orders in the m$p$-m$h$ expansion should be
necessary to describe ground state correlations compared to the
case without single-particle state renormalization.
An approximate but numerically far less demanding way of
implementing such a self-consistent definition of single-particle states
is to derive them from the mean-field calculated with the
{\it correlated} ground state one-body density matrix. This latter kind of
self-consistency has been tested in Ref.\cite{Pillet02} for
correlations generated by a zero-range residual interaction in A$\simeq$180
nuclei.
The result was that the m$p$-m$h$ expansion could be truncated at
4 particles--4 holes.

Taking the single-particle states entering the m$p$-m$h$ expansion
from a self-consistent mean-field is the essential difference between the
method of Ref.\cite{Pillet02} and the particle number conserving treatments
of the pairing
hamiltonian proposed in Ref.\cite{Wu89+,Molique}. It differs
in the same respect from the large-scale shell model approach which is widely
applied to light
nuclei and to heavier nuclei near closed
shells~\cite{shellmod}. In addition, contrary to
shell model calculations, the whole single-particle spectrum is {\it a priori}
considered in the definition of m$p$-m$h$ configurations, the space truncation
resulting from imposing a limit on the
order $\alpha$ of the
$\alpha p$-$\alpha h$ configurations included.

In principle, given a many-body hamiltonian capable of describing both the
nuclear mean-field and the main kinds of correlations beyond the mean field,
such an approach can be implemented in a fully microscopic way. This can be
done in particular by choosing a many-body effective hamiltonian built with the
standard density dependent effective forces which have proved to
give a precise account of pairing and RPA correlations.
In this case, the coefficients of the m$p$-m$h$ expansion can be derived by
applying a variational principle to the total nuclear energy.
The application of the method of Ref.\cite{Pillet02} then appears as a
straightforward
generalization of HFB and RPA calculations, without
the above mentioned particle number and Pauli violations.

Still, conceivable applications to realistic situations in nuclei require that
one is able to describe relevant correlations using a
small enough number
of particle-hole excitations
in the m$p$-m$h$ expansion.
In order to test the capacity of the method to describe the
correlations generated by a pairing residual interaction, we have applied it
to the exactly solvable hamiltonian proposed
long time ago by Richardson~\cite{Richard64}.
This simple model consists of a system of 2N identical
fermions distributed over a set of 2N equally spaced, twofold degenerate levels
\cite{Richard66}. The m$p$-m$h$ method has been applied to this hamiltonian
under the form of
multiple $p\bar{p}$-$h\bar{h}$ excitations,
where $\bar{p}$ and $\bar{h}$ are the time-reversed
of $p$ and $h$.
Particle numbers and pairing strengths have been explored within ranges of
values corresponding to typical situations found in realistic nuclei. Ground
state correlation energies, occupation probabilities and odd-even mass
differences have been calculated and compared with the exact results given by
Richardson
for different numbers of $p\bar{p}$-$h\bar{h}$ excitations.
Let us emphasize that the purpose of the present work is not to discuss
the interpretation of the calculated quantities
and their connection to pairing in real nuclei. Our goal is simply to
check the ability of the  m$p$-m$h$ approach to describe pairing correlations
within a model where exact solutions exist.

The paper is organized as follows. In Section II we present additional detail
concerning the m$p$-m$h$ configuration mixing method. We also give the
Richardson expression of the exact ground state solution of the pairing
hamiltonian. Results obtained with the m$p$-m$h$ method
are presented and discussed in Section III. A summary and conclusions are given
in Section IV.

\section{The Formalism}

In the m$p$-m$h$ configuration mixing method the ground state wave function of
the system is written as a superposition of a HF Slater determinant
$\dspt|\phi_{0}>=\prod_{h}a^+_h |0>$ and particle-hole
excitations built upon it:
\begin{equation}
\begin{array}{rl}
|\Psi>=& \dspt A_{0p0h}|\phi_0>+ \sum_{i}A_{1p1h}^{i}
                                a_{p_i}^{+}a_{h_i}|\phi_0> + \\
  &  \dspt \frac{1}{2!}\sum_{i,j}A_{2p2h}^{ij}
a_{p_j}^{+}a_{h_j} a_{p_i}^{+} a_{h_i}   |\phi_0> + \ldots
\end{array}
\label{e1}\end{equation}
The creation operators associated with unoccupied and occupied single particle
states in $|\phi_{0}>$ are denoted by $a^+_p$ and $a^+_h$, respectively
and the $A_{\alpha p\alpha h}$ are configuration mixing coefficients to be
determined. In compact form, eq.(\ref{e1}) can be written:
\begin{equation}
|\Psi> = \dspt \sum_{\alpha =0}^{M} \sum_{{\bf i}=1}^{m_{\alpha}}
A_{\alpha p \alpha h}^{{\bf i}} |\phi_{\alpha p_{{\bf i}} \alpha
h_{{\bf i}}}> ,
\label{e2}\end{equation}
where ${\bf i}$ denotes an ordered set of indices specifying a given $\alpha
p$-$\alpha h$ configuration, $|\phi_{\alpha p_{{\bf i}} \alpha h_{{\bf i}}}>$
is the wave function obtained by acting with the corresponding
$\alpha p$-$\alpha h$ operator on $|\phi_{0}> \equiv
|\phi_{0p0h}>$, $m_{\alpha}$ is the number of configurations of
$\alpha p$-$\alpha h $ type ($m_{0}=1$) and M is an integer parameter.
The configuration mixing coefficients $A_{\alpha p
\alpha h}^{{\bf i}}$ are obtained by minimizing the total energy
of the system:
\begin{equation}\dspt
\dspt \frac{\partial <\Psi| {\hat{H} - \lambda} |\Psi>}
{\partial A_{\alpha p \alpha h}^{{\bf i}~*}} =
 \frac{\partial <\Psi| {\hat{H} - \lambda}
|\Psi>}{\partial A_{\alpha p \alpha h}^{{\bf i}}} = 0 ,
\label{e3}\end{equation}
where $\lambda$ is the Lagrange parameter related to the
conservation of the norm of $|\Psi>$.

When $\hat{H}$ is independent of the nuclear density, the conditions
(\ref{e3}) are equivalent to the secular equation expressing
the diagonalization of $\hat{H}$ in the m$p$-m$h$ space :
\begin{equation}
\sum_{\alpha'} \sum_{{\bf i'}} {<\phi_{\alpha_{p_{{\bf i}}}
\alpha_{h_{{\bf i}}}}|\hat{H}|
\phi_{\alpha'_{p_{{\bf i'}}}\alpha'_{h_{{\bf i'}}}}>}
A_{\alpha' p \alpha' h}^{{\bf i'}}
 = \lambda  \; A_{\alpha p \alpha h}^{{\bf i}}
\label{e4}\end{equation}
and the complex conjugate one.

In this work we choose for $\hat{H}$ the pairing hamiltonian:
\begin{equation}
\hat{H} = \dspt \sum_{f=1}^{2N} \epsilon_{f} (a_{f}^{+}
a_{f}+a_{\bar{f}}^{+} a_{\bar{f}})
-g \sum_{f=1}^{2N}\sum_{f'=1}^{2N} a_{f}^{+}a_{\bar{f}}^{+} a_{\bar{f'}} a_{f'} ,
\label{e5}\end{equation}
where ${\bar{f}}$ denotes the time-reversed state of $f$.
The exact ground state of this hamiltonian for the case
of a system of 2N fermions has been derived by Richardson~\cite{Richard66}:
\begin{equation}
|\Psi^{exact}>=\prod_{i=1}^{N}B_{i}^{+}|0> ,
\label{e6}\end{equation}
where the operator $B_i^+$ creates a collective pair :
\begin{equation}
B_{i}^{+}=\sum_{j=1}^{N} \frac{1}{2\epsilon_j-E_i} a_j^+ a_{\bar{j}}^{+} ,
\label{e7}\end{equation}
and $|0>$ is the fermion vacuum.
The quantities $E_i$ are the solutions of a set
of N nonlinear equations \cite{Richard66} :
\begin{equation}
1-2g \sum_{j(\neq i)=1}^{N} \frac {1} {E_j-E_i} +
g \sum_{j=1}^{N} \frac{1} {2\epsilon_j-E_i}=0 .
\label{e8}\end{equation}
and the exact ground state energy of the system is given by:
\begin{equation}
E= \sum_{i=1}^{N}E_i .
\label{e9}\end{equation}

The solutions of the pairing hamiltonian can be classified by the
eigenvalues of the seniority operator:
\begin{equation}
\dspt \hat{\nu} = \sum_{f}  (a_{f}^{+}a_{f} -a_{\bar{f}}^{+}a_{\bar{f}})^2 ,
\label{e10}\end{equation}
which counts the number of unpaired particles present in the system.
In the m$p$-m$h$ scheme the ground state of the pairing hamiltonian
which corresponds to seniority zero can be generated by
considering in (\ref{e2}) only configurations
$\alpha p$-$\alpha h=\beta p\bar{p}$-$\beta h\bar{h}$
with even values $\alpha$=2$\beta$.
If one includes in expansion (\ref{e2}) all $\alpha p$-$\alpha h$
configurations up to $M$=2$N$, the wave function $|\Psi>$
becomes equivalent to the exact wave function $|\Psi^{exact}>$.

It is interesting to discuss the relationship between the m$p$-m$h$
configuration mixing method presented above and the projected-BCS
(PBCS) approach \cite{projection,pbcs},
employed for restoring
particle number conservation in finite Fermi systems.
In the PBCS approach the ground state has a pair condensate
structure obtained by replacing in eq.(\ref{e6}) the  N pair operators
$B_i^+$  by a unique collective pair :
\begin{equation}{\Gamma^{+}=\sum_{j=1}^{N} x_j a_j^+ a_{\bar{j}}^{+} ,
}\label{e11}
\end{equation}
where the amplitudes $x_j$ are determined variationally.
The PBCS wave function can be written also in terms of particle-hole
operators acting on the HF state \cite{ds}:
\begin{equation}{
|PBCS> = const. \sum_{n=0} \frac {(\Gamma_p^+ \Gamma_h)^n}{(n!)^2}
|HF> , }\label{e12}
\end{equation}
where the pair operators $\Gamma_h$ ($\Gamma_p$) are formed by
restricting the summation in (\ref{e11}) to hole (particle)
states and replacing $x_j$ by $x_p$ ($1/x_h$).
From (\ref{e12}) it can be seen that the PBCS wave
function belongs to a particular subset of the m$p$-m$h$ wave functions
(\ref{e2}) in which the mixing coefficients have a separable form in
the indices associated with particles and holes.
It is known that, when the $x_j$'s are determined from projection before
variation, PBCS wave-functions give a satisfactory description of
weak pairing regimes and of the crossover from normal to superfluid
phases in Fermi systems. One therefore expects that the more general
m$p$-m$h$ form (\ref{e1}) is able to provide an accurate
description of pairing correlations in these situations. In addition, since the
$x_p$ and $1/x_h$ are small, only the first few terms
in (\ref{e1}) should be necessary.

In order to test the convergence of expansion (\ref{e1}) in all pairing
regimes, the ground state solution of the hamiltonian (\ref{e6}) has been
calculated using eq.(\ref{e4}) for two systems, one with
2N=8 particles distributed among 8 equidistant levels, and the
other one with 2N=16 particles distributed among 16 equidistant
levels. The levels are twofold degenerate.
These two systems simulate the level densities and pairing diffusivities found
in weakly and moderately superfluid nuclei, respectively.
The constant $g$ has been varied in a large range of values around typical
pairing strengths in nuclei.

\section{Numerical Results}

The number of seniority-zero states of the 2N particle
system is equal to the binomial coefficient $C_{2N}^{N}$. Thus,
for the systems with 2N=8 and 2N=16 particles the total number of
seniority-zero states which should eventually be used for
exactly calculating the ground states of the two systems is equal to
70 and 12870, respectively. We study here the convergence
properties of the m$p$-m$h$ expansion
(\ref{e1}) as a function of
the maximum particle-hole order $M$
and of an energy cut-off in the energy of the m$p$-m$h$ configurations.
The numbers of $\alpha p\bar{p}$-$\alpha h\bar{h}$
configurations for
$2\alpha$ between 2 and 8 are shown in Table~\ref{table1}
for the two systems.
The highest number of configurations is reached when
$\alpha$ is equal to half the number of particles.
The numbers of configurations for various
cut-off energies $E_{cut}$ are given in Table~\ref{table2}.
Here and in the  following, energies are in units of the equidistant level
spacing $d$.
\begin{table}
\begin{center}
\begin{tabular}{|c||c|c|c|c|}
\hline
$ $ 2N     &   2p2h      &  4p4h      &  6p6h      &   8p8h   \\
\hline
\hline
$   8  $   &    16       &  36        &  16        &     1   \\
\hline
$  16  $   &    64       &  784       &  3136      &    4900  \\
\hline
\end{tabular}
\end{center}
\caption{
Number of $\alpha p\bar{p}$-$\alpha h\bar{h}$ configurations
for given
values of $2\alpha$ between 2 and 8 for the
systems with 2N=8 and 2N=16 particles.}
\label{table1}\end{table}
\begin{table}
\begin{center}
\begin{tabular}{|c||c|c|c|c||c||c|c|c|c|}
\hline
$E_{cut}$ & 2p2h & 4p4h & 6p6h & 8p8h & $E_{cut}$ & 2p2h & 4p4h & 6p6h & 8p8h \\
\hline \hline
 2     & 1  & 0 & 0 &  0 & 2     & 1  & 0 & 0 &  0  \\
\hline
 8     & 10  & 1 & 0 &  0 & 8     & 10  & 1 & 0 &  0 \\
\hline
 18    & 16  & 28 & 1 &  0 & 18    & 43  & 50 & 1 &  0 \\
\hline
 32    & 16  & 36 & 16 &  1 & 32    & 64  & 428 & 181 &  1 \\
 \hline
   &  &  &  &  & 50      &  64 &    776 &  1946 &   464    \\
\hline
   &  &  &  &  & 70      &  64 &    784 &  3118 &  3710    \\
\hline
   &  &  &  &  & 128    &   64 &    784 &  3136 &  4862    \\
\hline
\end{tabular}
\end{center}
\caption{Number of configurations corresponding to various cut-off
energies $E_{cut}$ for a system of 2N particles. Left: 2N=8.
Right: 2N=16. $E_{cut}$ is in unit of the level spacing $d$.}
\label{table2}\end{table}

The convergence of the m$p$-m$h$ expansion towards the exact
solution is studied below by looking at correlation energies, odd-even mass
differences and level occupation probabilities.
The same quantities calculated with the plain BCS (unprojected) approach are
also given.

\subsection{Ground state correlation energy}

The correlation energy $E_{corr}$ in the ground state of the
system is taken as:
\begin{equation}
E_{corr} = E(g \neq 0) - E(g=0) ,
\label{e13}\end{equation} where $E(g)$ is the
total energy of the interacting system.

The amount of pairing correlations
in the 2N particle
ground state
can be
estimated from
the odd-even mass difference defined by \cite{rs} :
\begin{equation}
P(2N)=2E(2N-1)-E(2N)-E(2N-2) ,
\label{e14}\end{equation}
where
$E(N)$ is the ground state energy for
$N$ fermions.
In the calculations we adopted the convention of Ref.\cite{Richard66}, i.e.,
for a system of
$N$ particles
($N$ odd or even),
the pairing force is effective among
$N$ doubly-degenerated levels.
Notice that $P(2N)=0$ for non-interacting particles ($g=0$).
Therefore, the mean-field effect which is contained in odd-even mass
differences for real nuclei~\cite{satula} is not present in the equidistant
model used here.
Thus, $P(2N)$ can be taken as a genuine measure of the
intensity of pairing correlations.

Assimilating $P(2N)$ with the pairing gap $\Delta$, the
weak (strong) pairing regime corresponds to values much smaller (larger)
than 1 of the ratio $\eta=\Delta/d$.
For deformed nuclei in the rare earth region $\Delta$ is
about 1 MeV while $d$ is of the order of 400-500 keV.
Thus, for these nuclei $\eta$ is of the order of 2-3. As we shall see
below, for values of $\eta$ in this range the correlation energies
evaluated in the BCS approximation are largely underestimated
while the m$p$-m$h$ expansion converges quickly to the exact
results.
Let us add that, in units of $d$, the constant $g$ corresponding to
typical nuclear pairing strengths is about 0.5.

\begin{figure}[h]
\begin{center}
\vspace{5mm}
\includegraphics[height=5.5cm,angle=0]{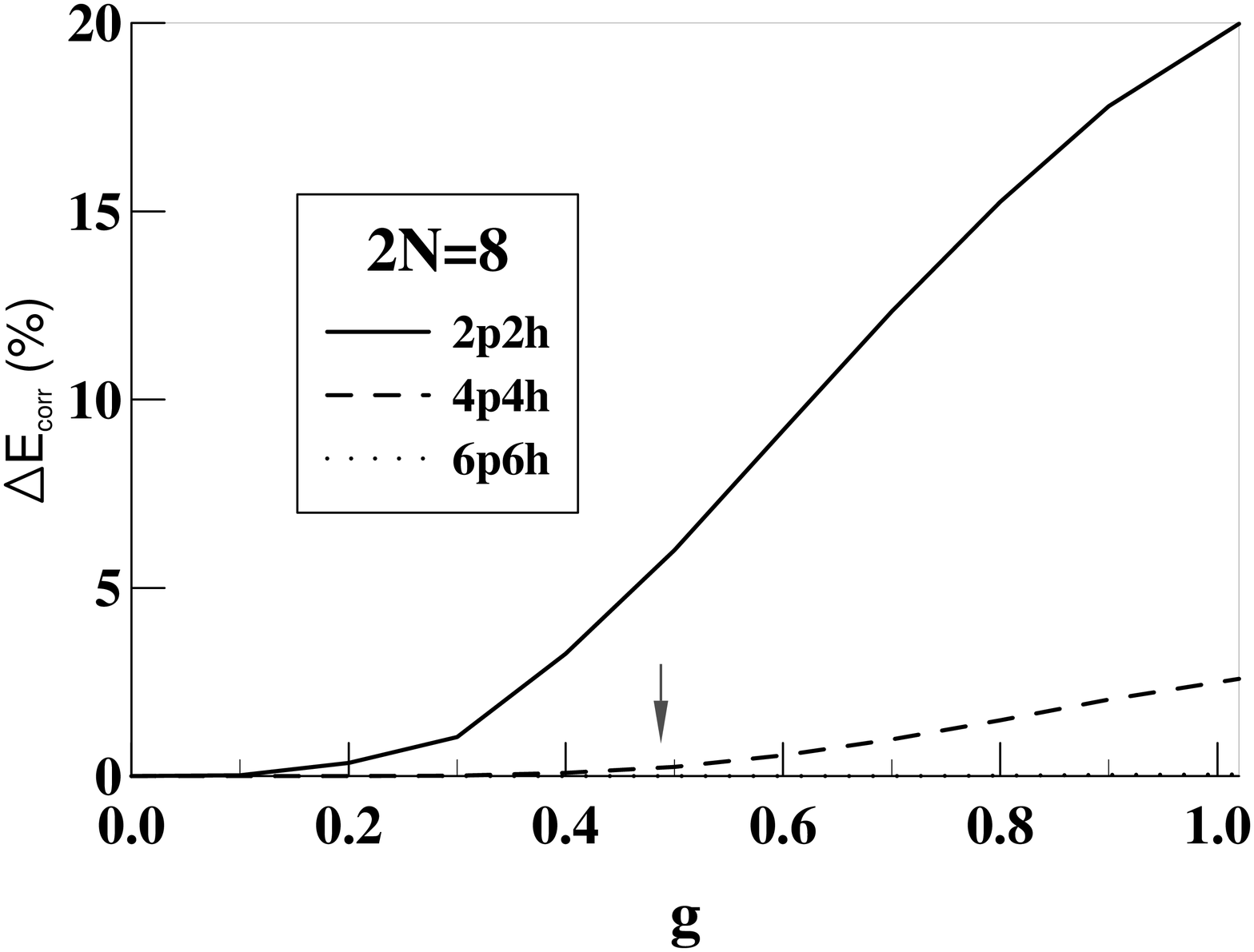}
\vspace{-5mm}
\includegraphics*[height=5.5cm,angle=0]{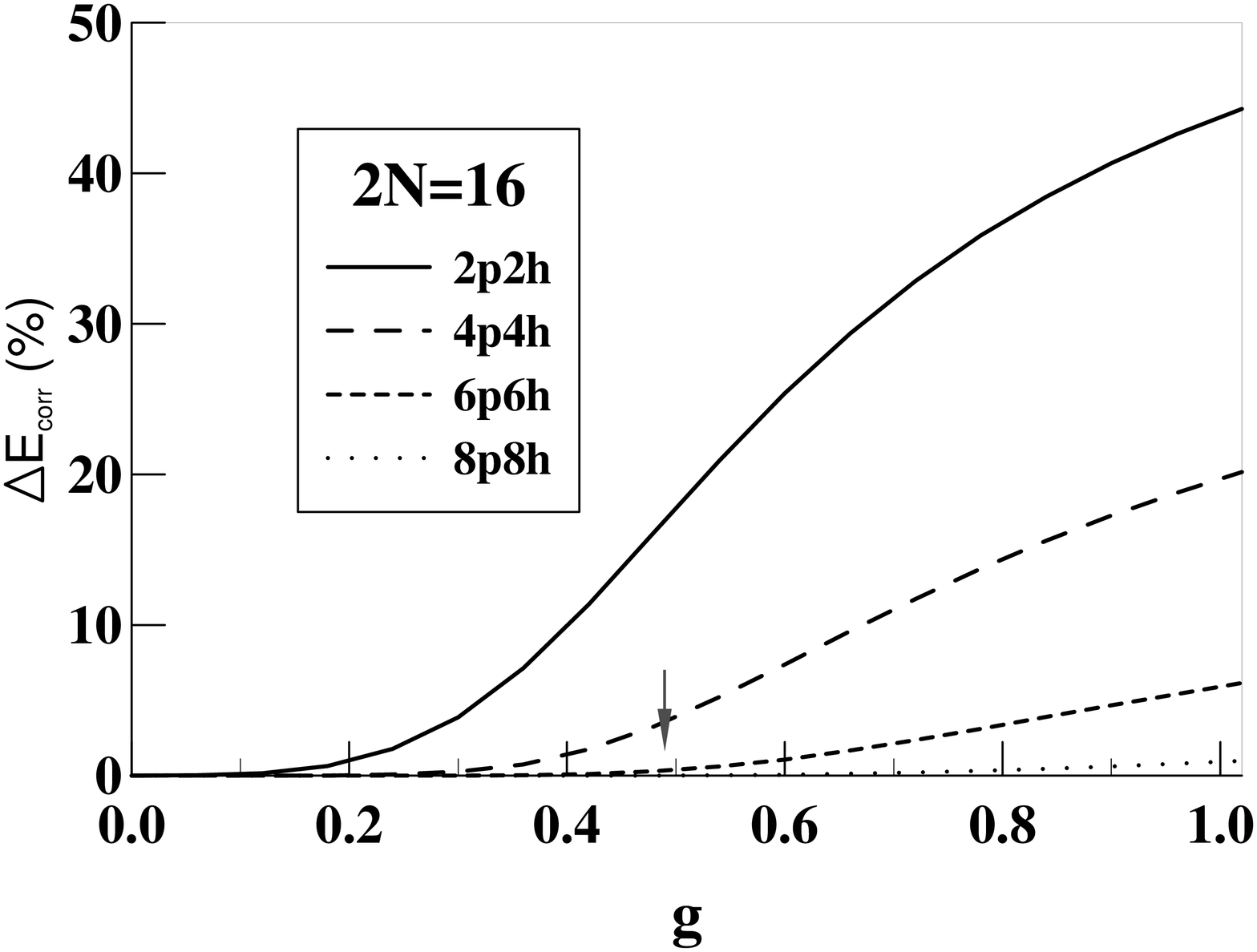}
\end{center}
\caption{Percentage error on correlation energies as a function of
the pairing strength $g$ for the 2N=8 and 2N=16 cases.
The curves correspond to the results obtained by
cutting the m$p$-m$h$ expansion at 2p2h, 4p4h and 6p6h.}
\label{figure1}
\end{figure}

We first analyze the convergence of the m$p$-m$h$ expansion for the
two quantities $E_{corr}$ and $P(2N)$
when including successively all
$2p2h$, $4p4h$,~\ldots, configurations
in the ground state wave function (2)
(without any cut-off).
The results of the calculations
for the systems with 8 and 16 particles are shown in
Tables~\ref{table3}-\ref{table4} and in Fig.\ref{figure1}.
The errors for correlation energies,
$\Delta E_{corr}$, and for the odd-even mass differences, $\Delta P$, are
calculated relative to the exact values. As mentioned above, for
the system with 2N=8 particles an expansion up to $8p-8h$
corresponds to the exact solution.

\begin{table}
\begin{center}
\begin{tabular}{|c||c|c|c|c|c|}
\hline
    g       &  npnh      &  $E_{corr}$      & $\Delta E_{corr}(\%)$ &   $P(2N)$  & $\Delta P(2N)(\%)$ \\
\hline
\hline
    0.2    &  2         &    -0.924       &   0.36          &  0.299     &     1.65   \\
           &  4         &    -0.927       &   0.00          &  0.303     &     0.33   \\
           &  6         &    -0.927       &   0.00          &  0.304     &     0.00   \\
           &  8         &    -0.927       &   0.00          &  0.304     &     0.00   \\
           &  BCS       &       0         &   100.          &   0        &     100.   \\
           &  Exact     &    -0.927       &                  &  0.304     &        \\
\hline
    0.6    &  2         &    -3.757       &   9.18          &  1.381     &     20.59  \\
           &  4         &    -4.114       &   0.55          &  1.718     &      1.27  \\
           &  6         &    -4.136       &   0.01          &  1.739     &      0.00  \\
           &  8         &    -4.136       &   0.00          &  1.739     &      0.00  \\
           &  BCS       &    -2.942       &  28.92          &  1.570     &     9.72   \\
           &  Exact     &    -4.136       &                 &  1.739     &            \\
\hline
    0.8    &  2         &    -5.602       &  15.25          &  2.618     &     29.21      \\
           &  4         &    -6.511       &   1.49          &  3.594     &      2.81      \\
           &  6         &    -6.608       &   0.02          &  3.696     &      0.05      \\
           &  8         &    -6.610       &   0.00          &  3.701     &      0.00      \\
           &  BCS       &    -5.028       &  23.93          &  2.502     &     32.40      \\
           &  Exact     &    -6.610       &                 &  3.701     &                \\
\hline
\end{tabular}
\end{center}
\caption{Correlation energies $E_{corr}$ and odd-even mass
differences $P(2N)$ for a system with 2N=8 particles. The results
are shown for several values of
the pairing strength $g$ and
with the m$p$-m$h$ expansion cut at various orders.
The BCS and exact results are also shown. In the BCS case,
the value given in the column P(2N) is the one
of the pairing gap.}
\label{table3}
\end{table}

\begin{table}
\begin{center}
\begin{tabular}{|c||c|c|c|c|c|}
\hline
    g       &  npnh      &  $E_{corr}$      & $\Delta E_{corr}(\%)$ &   $P(2N)$  & $\Delta P(2N)(\%)$ \\
\hline
\hline
    0.18   &  2         &    -1.659       &  0.64          & 0.305      &  4.09      \\
           &  4         &    -1.669       &  0.01          & 0.317      &  0.31      \\
           &  6         &    -1.669       &  0.00          & 0.318      &  0.00      \\
           &  8         &    -1.669       &  0.00          & 0.318      &  0.00      \\
           &  BCS       &     0.00        &  100.00    &     0.00       & 100.00       \\
           &  Exact     &    -1.669       &                & 0.318      &            \\
\hline
    0.54   &  2         &    -6.883       &  20.92         &  1.941     &     49.51     \\
           &  4         &    -8.250       &   5.22         &  3.301     &     14.13     \\
           &  6         &    -8.652       &   0.60         &  3.781     &      1.64     \\
           &  8         &    -8.702       &   0.03         &  3.842     &      0.05     \\
           &  BCS       &    -6.636       &  23.76         &  2.580     &   32.88    \\
           &  Exact     &    -8.704       &                &  3.844     &           \\
\hline
    0.66   &  2         &    -9.149       &  29.37         &  2.598     &     57.16      \\
           &  4         &    -11.711      &   9.59         &  4.746     &     21.74      \\
           &  6         &    -12.738      &   1.66         &  5.819     &     4.04     \\
           &  8         &    -12.938      &   0.12         &  6.047     &     0.28      \\
           &  BCS       &    -10.357       &   20.05        &  3.701     &   43.96     \\
           &  Exact     &    -12.954      &                &  6.064     &           \\
\hline
\end{tabular}
\end{center}
\caption{Same as Table~\ref{table3}, for 2N=16 particles.}
\label{table4}
\end{table}

In Tables~\ref{table3}-\ref{table4} the results are shown for several values
of the pairing strength $g$. In the BCS approximation there is a sharp
transition between the normal and the superfluid phase which
appears at $g_c \approx 0.31$ for 2N=8 and $g_c \approx 0.24$ for
2N=16. Below these critical values of the pairing strength the
correlation energies are zero in the BCS approximation. This is
not the case for the exact solution, which shows a rather smooth
increase of the correlation energy
with the pairing strength.

From Tables~\ref{table3}-\ref{table4} we also notice that, in the BCS
approximation,
the correlation energies are strongly underestimated. The error made
is more than $20\%$ for all  the values of the pairing strength listed
in Tables~\ref{table3}-\ref{table4}. The same large differences are obtained
between the pairing gaps associated in BCS with the odd-even mass differences,
and the exact values of P(2N). It is worth stressing that the errors
in the BCS correlation energies remain rather large,  more than $10\%$,
even for large  values of the pairing strength for which the
ratio $\eta$ between the BCS pairing gap and the level spacing
(which in our case is the energy unit) is of the order of 20.

Let us emphasize however that in BCS or HFB calculations applied to
real nuclei, the pairing strengths for neutrons and protons -- or the
matrix elements of the pairing residual interaction -- are adjusted in
order to correctly describe pairing correlations in strongly superfluid
nuclei. Consequently, BCS or HFB results should be taken with caution
only in weak pairing regimes.

From Tables~\ref{table3}-\ref{table4} one can see that the
m$p$-m$h$ expansion is converging rapidly to the exact results.
Thus, for the system with 16 particles a truncation to $6p$-$6h$
provides rather accurate values for the correlation energies
$(\Delta E_{corr} \sim 1 \%)$. This truncation order should be
compared to the maximum possible order of the m$p$-m$h$ expansion,
which is $16p$-$16h$ in this case. As seen in Table~\ref{table1}, for a
truncation at $6p$-$6h$ the number of configurations is equal to
3136, while the total number of seniority-zero configurations is
12870.

\begin{figure}[h]
\includegraphics*[height=6.5cm,angle=0]{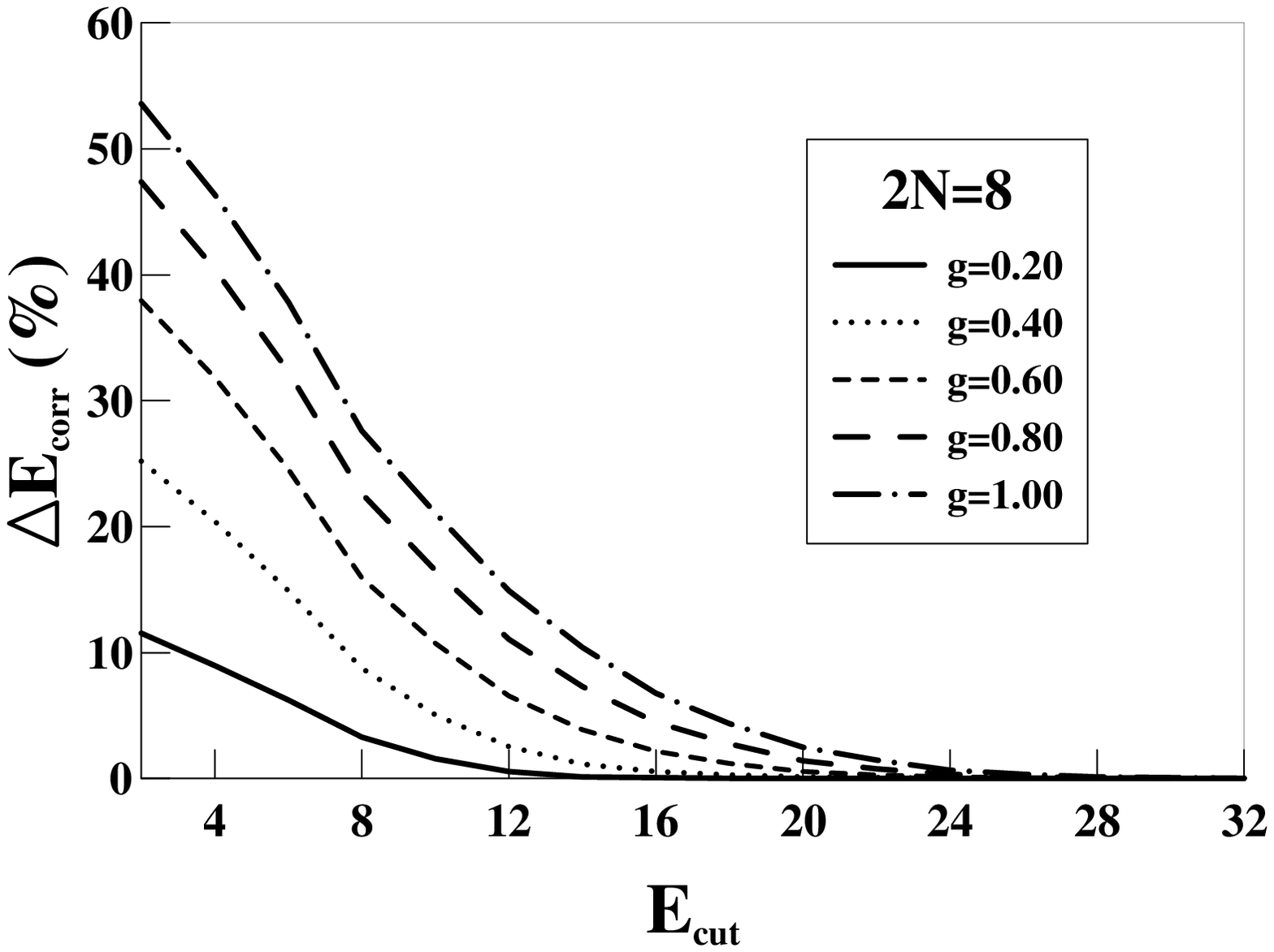}
\includegraphics*[ height=6.5cm,angle=0]{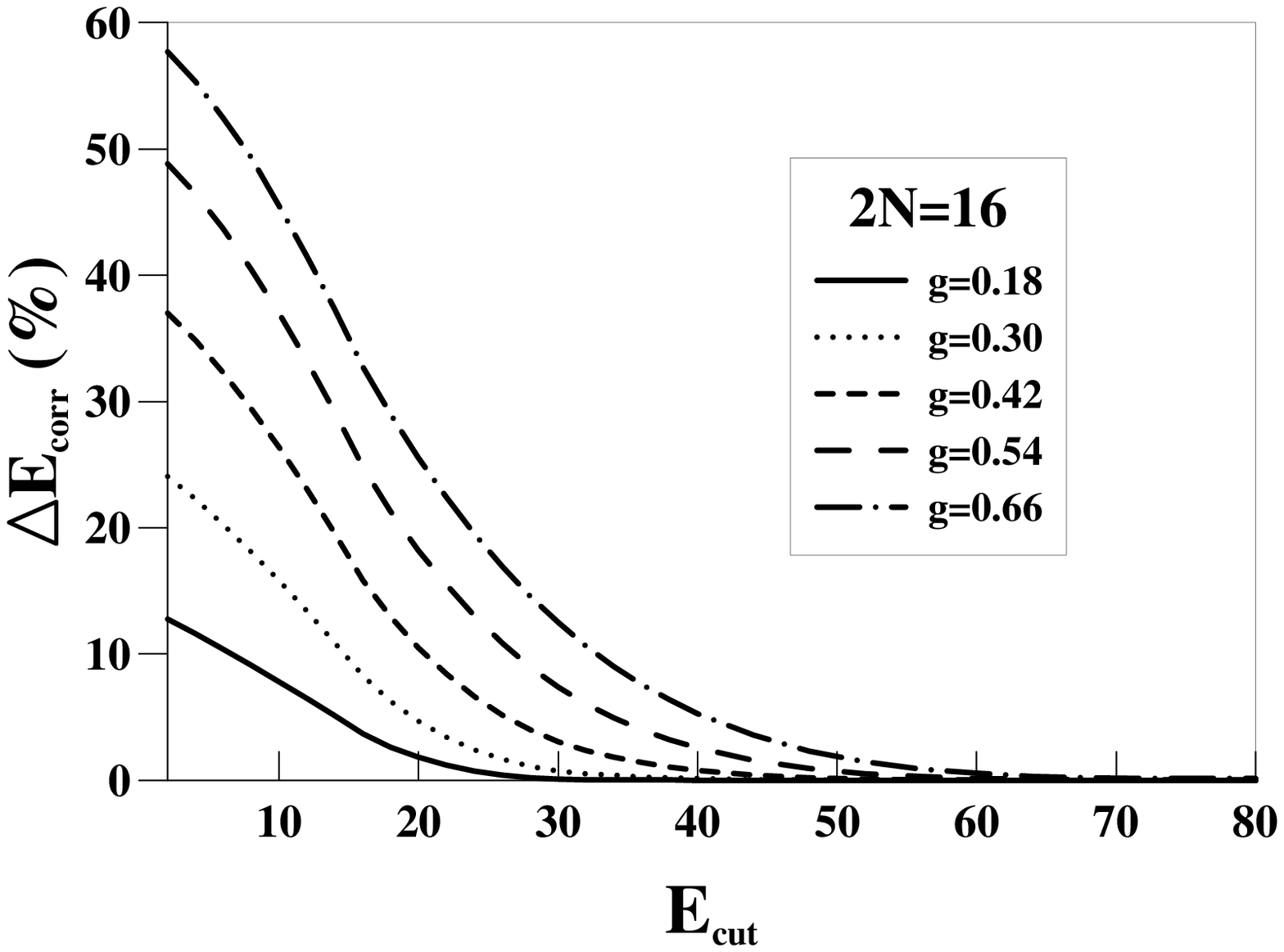}
\hspace{-10mm}
\caption{Percentage errors on correlation energies as a function
of cut-off energy $E_{cut}$. The curves corresponds to various
pairing strength values listed in the inset. The results are for
the systems with 2N=8 and 2N=16 particles.}
\label{figure2}\end{figure}

Next, we examine the convergence of the results as a function of
the energy cut-off $E_{cut}$. In this truncation scheme we
consider only m$p$-m$h$ configurations whose excitation energies are
smaller than $E_{cut}$. The number of
configurations for various $E_{cut}$ values are shown in Table~\ref{table2}.
In Fig.\ref{figure2} the evolution of $\Delta E_{corr}$
with $E_{cut}$ is shown for several values of the pairing strength $g$.

\begin{figure}[h]
\begin{minipage}[t]{75mm}
\includegraphics[width=7.5cm,scale=1.]{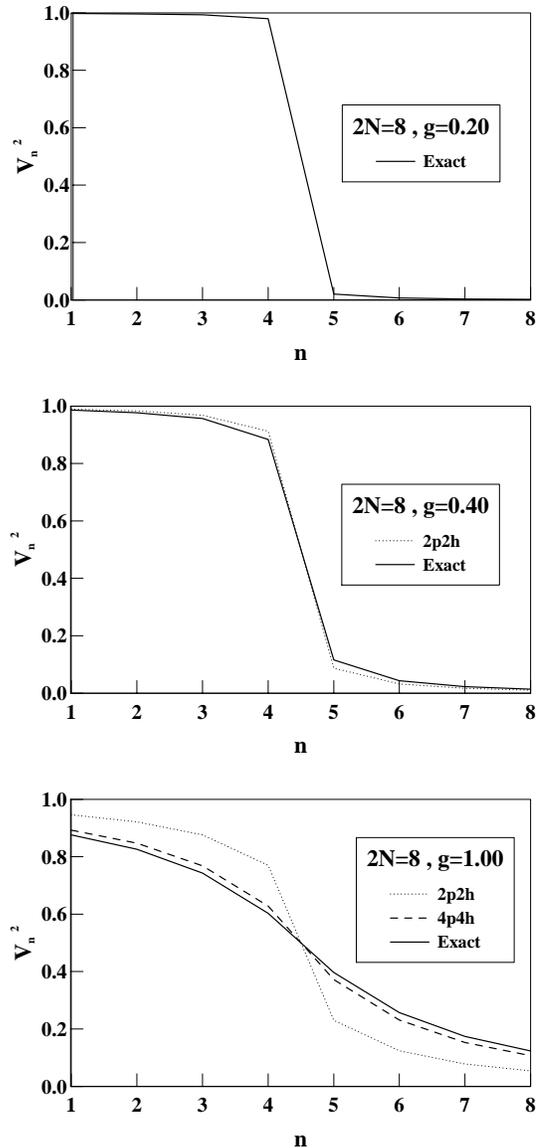}
\vspace{-7mm}
\caption{ Occupation probabilities $v_n^2$ for the single-particle
levels $n$. The calculated points
are joined by straight lines. The curves correspond to different
truncation orders specified in the insets, the full line showing
the exact values. The results are for the system with 2N=8
particles.}
\label{figure5}
\end{minipage}
\end{figure}

From these
figures and from Table~\ref{table2} one can see that, for the values of $g$
of physical interest ($g \sim 0.5$) the number of configurations needed
to achieve the same accuracy as previously is smaller.
Thus, in the case of the system with 16 particles and for
g=0.54 one needs a cut-off energy $E_{cut}$=50 in order to get an
accuracy of $\Delta E_{corr} \sim 0.5\%$. As seen in Table~\ref{table2}, for
this cut-off energy one selects 3250 configurations up to 8$p$-8$h$.
On the other hand, in the truncation scheme based on the order of the
particle-hole expansion, to achieve the same accuracy one needed to
consider all configurations up to $6p$-$6h$ that is, from Table~\ref{table1},
3984 configurations. This is about $20\%$ more than in the truncation
based on  cut-off energy.
In particular a large number --
about one third -- of the 6$p$-6$h$ configurations of Table~\ref{table1}
are eliminated by the cut-off. Since the same precision is achieved, this
means that these eliminated configurations do not contribute to the correlation
energy. Let us mention that a 50 MeV cut-off in the excitation energy of
scattered pairs is consistent with the results
of BCS or HFB calculations in actual nuclei, even in strongly
superfluid ones.

\begin{figure}
\begin{minipage}[t]{75mm}
\vspace{-5mm}
\includegraphics[width=7.5cm,scale=1.]{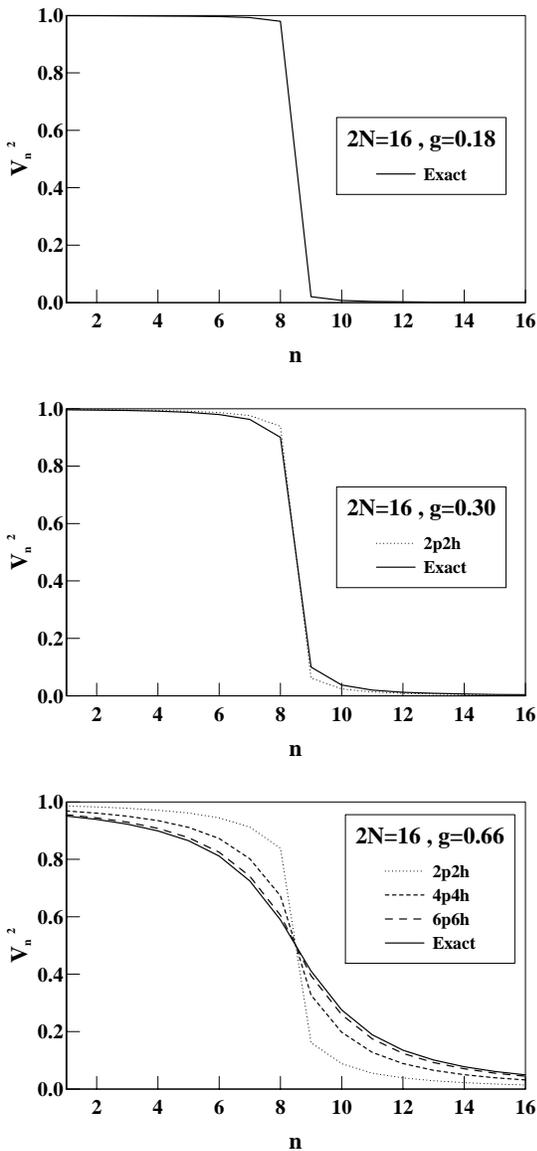}
\vspace{-8mm}
\caption{ Same as Fig.\ref{figure5} for 2N=16 particles.}
\label{figure6}
\end{minipage}
\end{figure}

\subsection{Occupation probabilities}

The occupation probability $v_n^2$ of the single-particle level
$n$ is given by :

\begin{equation}
v_{n}^{2}=\frac {1}{2} <\Psi| a_n^+ a_n + a_{\bar{n}}^+
a_{\bar{n}}|\Psi> .
\label{e15}\end{equation}

The occupation probabilities of all single-particle levels in the cases 2N=8
and 2N=16 are plotted in Figs.\ref{figure5}-\ref{figure6} for
three values of the pairing strength $g$. It can be seen that for
small values of $g$ the exact values of occupation probabilities
are rather well reproduced by a truncation to $2p2h$
configurations.

However, for the highest values of $g$ listed in
Tables~\ref{table3}-\ref{table4} i.e., $g$=1.0 for 2N=8  and $g$=0.66
for 2N=16, one needs to introduce up to $6p$-$6h$ configurations
in order to get an accurate description of occupation
probabilities in the m$p$-m$h$ scheme.
This truncation order appears consistent with
the behaviour of correlation energies discussed in the previous
subsection.

\section{Summary and Conclusions}

In this paper we have analyzed the convergence properties of the
multiparticle-multihole (m$p$-m$h$) configuration mixing approach.
The study was done for a pairing hamiltonian with a constant
pairing force, which can be solved exactly. For the
single-particle model we have chosen a sequence of
doubly-degenerate and equidistant levels, half-filled with one
kind of fermions. The parameters of the model and the number of
particles have been chosen so as to simulate the physical situation met
in light and medium nuclei.
The pairing strength $g$ has been varied from zero up to about
twice the value corresponding to typical conditions in atomic nuclei.

We have shown that a truncation
based on the m$p$-m$h$ expansion is converging rather rapidly to
the exact results for the ground state correlation energies and
occupation probabilities of single-particle levels. For instance,
with a value of $g$ corresponding to standard nuclear pairing,
one can get an accuracy of about
$1\%$ for the correlation energy in a system of 16 particles
if one considers configurations
up to $6p$-$6h$. The number of these configurations is about four
times smaller than the total number of possible m$p$-m$h$
configurations with seniority zero that the system can form.
An even faster convergence without precision loss is found by selecting
configurations whose excitation energies are limited by a cut-off
corresponding to the maximum energy of scattered pairs.
These encouraging results indicate that applications of the m$p$-m$h$ approach
to fully microscopic self-consistent calculations in
nuclei should be possible with present-day numerical capabilities.

\vskip 0.05cm \noindent {\bf Acknowledgments}: We thank J.
Dukelsky for providing us with his computer code for solving the Richardson
equations.

\end{document}